\documentclass[12pt,a4paper]{article}
\usepackage{epsfig}
\begin{document}
\title{Lepton flavor violating signals of gauge bosons $Z^{\prime}$ \\
at future $e^{+}e^{-}$ colliders }
\author{Chongxing Yue$^{a}$, Yanming Zhang$^{b}$, Hong Li$^{b}$\\
{\small a:Department of Physics, Liaoning Normal University,
Dalian 116029 P. R. China}\thanks{E-mail:cxyue@lnnu.edu.cn}
\thanks{This work is supported by the National Natural Science
Foundation of China(90203005)}
\\ {\small b:College of Physics and Information Engineering,}\\
\small{Henan Normal University, Xinxiang  453002 P. R.  China}
  }
\date{\today}
\maketitle
\begin{abstract}
\hspace{5mm} Many models, such as  topcolor-assisted technicolor
(TC2) models and flavor-universal TC2 models, predict the
existence of extra $U(1)$ gauge bosons $Z^{\prime}$, which couple
preferentially to the third generation. The lepton flavor
violating (LFV) signals of these new particles at the future
$e^{+}e^{-}$ colliders (LCs) are discussed via calculating its
contributions to the LFV process $e^{+}e^{-}\rightarrow\mu\tau$.
We calculate the cross section of this process and discuss its
dependence on the beam polarization. The ratio of signal over
square root of the background $(S/\sqrt{B})$ is also calculated.
The results show that the LFV signals of the gauge bosons
$Z^{\prime}$ may be detected at the future LC experiments.
\end {abstract}

\newpage
The high statistic results of the SuperKamiokande (SK) atmospheric
neutrino experiment \cite{y1} and the solar neutrino experiment
\cite{y2} have made one to believe that neutrinos are massive and
oscillate in flavor. This phenomenon is lepton flavor violating
(LFV). LFV processes are practically suppressed to zero in the
standard model(SM), due to the unitarity of he leptonic analog of
the CKM mixing matrix and the near masslessness of the three
neutrinos. So the experimental result is the first convincing
signature of new physics (NP) beyond the SM and make that one is
of interest in the LFV processes.

There are many kinds of NP scenarios predicting new particle,
which can lead to significant LFV signals. For example, in the
minimal supersymmetric SM, a large $\nu_{\mu}-\nu_{\tau}$ mixing
leads to clear LFV signals in slepton and sneutrino production and
in the decays of neutralinos and charginos into sleptons and
sneutrinos at hadron colliders and lepton colliders \cite{y3,y4}.
The non-universal $U(1)$ gauge bosons $Z^{\prime}$, which are
predicted by various specific models beyond the SM, can lead to
the large tree-level flavor changing (FC) couplings. Thus, these
new particles may have significant contributions to some LFV
processes \cite{y5,y6}.

Many kinds of models beyond the SM predict the presence of massive
 gauge bosons $Z^{\prime}$ that couple preferentially to the third generation
quarks and leptons. Examples are the topcolor-assisted technicolor
(TC2) models \cite{y7} and flavor-universal TC2 models \cite{y8}.
It has been shown that, to fit the electroweak measurement data,
the $Z^{\prime}$ mass must be larger than $1TeV$ \cite{y9}. Some
signals of this kind of extra gauge bosons have been studied in
literature \cite{y5,y6,y10}. In this paper, we are interested in
studying the contributions of $Z^{\prime}$ to the LFV process
$e^{+}e^{-}\rightarrow\mu\tau$, as a possible LFV signal of this
kind of gauge bosons at the future high energy $e^{+}e^{-}$ linear
collider (LC) experiments. We calculate the cross section of this
process and discuss its dependence on the beam polarization. The
main SM backgrounds of this process come from the Drell-Yan
process $e^{+}e^{-}\rightarrow
Z(\gamma)\rightarrow\tau^{+}\tau^{-}\rightarrow
\mu\bar{\nu}_{\mu}\tau\nu_{\tau}$ and $W^{+}W^{-}$ pair production
$e^{+}e^{-}\rightarrow W^{+}W^{-}\rightarrow
\mu\bar{\nu}_{\mu}\tau\bar{\nu}_{\tau}$. The ratio of signal over
square root of the background ($S/\sqrt{B}$) is simply calculated.
Our results show that there will be several tens and up to
thousand events to be generated at a future LC experiment with the
center-of-mass energy $\sqrt{s}=800GeV$ and a yearly integrated
luminosity of $L=580fb^{-1}$. The LFV signals of the extra
 $U(1)$ gauge bosons $Z^{\prime}$, which couple preferentially
 to the third generation, may be detected at the future LC experiments.

The key feature of TC2 models \cite{y7} and flavor-universal TC2
 models \cite{y8} is that the large top quark mass is
mainly generated by topcolor interactions at a scale of order
$1TeV$. The topcolor interactions may be flavor non-universal (as
in TC2 models) or flavor-universal (as in flavor-universal TC2
models). However, to tilt the chiral condensation in the
$t\bar{t}$ direction and not form a $b\bar{b}$ condensation, all
of these models need a non-universal extended hypercharge group
$U(1)$. Thus, the existence of the extra $U(1)$ gauge bosons
$Z^{\prime}$ is predicted. These new particles treat the third
generation fermions (quarks and leptons) differently from those in
the first and second generations, namely, couple preferentially to
the third generation fermions. After the mass diagonalization from
the flavor eigenbasis into the mass eigenbasis, these new
particles lead to tree-level FC couplings. The couplings of the
extra $U(1)$ gauge bosons $Z^{\prime}$ to ordinary fermions, which
are related to our calculation, can be written as \cite{y7,y11}:
\begin{equation}
{\cal L}=\frac{1}{2}g_{1}\{\tan\theta^{\prime}(\bar{e}_{L}\gamma^{\mu}
         e_{L}+2\bar{e}_{R}\gamma^{\mu}e_{R})-k_{\tau\mu}(\bar{\tau}_{L}
         \gamma^{\mu}\mu_{L}+2\bar{\tau}_{R}\gamma^{\mu}\mu_{R})\}\cdot Z'_{\mu},
\end{equation}
where $g_{1}$ is the hypercharge gauge coupling constant,
$\theta^{\prime}$ is the mixing angle with
$\tan\theta^{\prime}=\frac{g_{1}}{2\sqrt{\pi k_{1}}}$ and
$k_{\tau\mu}$ is flavor mixing factor. In the following estimation,
we will take $k_{\tau\mu}=\lambda$, where $\lambda=0.22$ is the
Wolfenstein parameter \cite{y12}.

 Since the electroweak gauge bosons $\gamma$ and $Z$ can not couple to
 $\mu\tau$ at tree-level,we need not consider the interference effects
 between the $\gamma$, $Z$ and $Z^{\prime}$ on the cross section of the process
$e^{+}e^{-}\rightarrow\mu\tau$. Then the total unpolarized cross section
$\sigma$ can be written as (at tree-level):
\begin{equation}
\sigma=\frac{25\pi^{2}\alpha^{3}k_{\tau\mu}^{2}}{12C_{W}^{6}k_{1}}\cdot
\frac{s}{(s-M_{Z}^{2})^{2}+M_{Z}^{2}\Gamma_{Z'}^{2}},
\end{equation}
where $C_{W}=\cos\theta_{W}$, $\theta_{W}$ is the Weinberg angle.
 $\sqrt{s}$ is the center-of-mass energy of LC experiments.
$M_{Z}$ is the mass of the extra $U(1)$ gauge boson $Z^{\prime}$
and $\Gamma_{Z'}$ is its total decay width. $\Gamma_{Z^{\prime}}$
is dominated by $t\bar{t}$ and $b\bar{b}$, which can be written as
\cite{y13}:
$$\Gamma_{Z'}\approx \frac{k_{1}M_{Z}}{3}$$

To obtain proper vacuum tilting (the topcolor interactions only
condense the top quark but not the bottom quark), the coupling
constant $k_{1}$ should satisfy certain constraint, i.e.
$k_{1}\leq 1$ \cite{y8}. The limits on the $Z^{\prime}$ mass
$M_{Z}$ can be obtained via studying its effects on experimental
observable \cite{y11}. Recently, Ref.[9] has shown that, to fit
the current electroweak measurement data, the $Z^{\prime}$ mass
must be larger than $1TeV$. As numerical estimation, we assume the
center-of-mass energy $\sqrt{s}=800GeV$ and take $k_{1}$, $M_{Z}$
as free parameters.

In Fig.1 we plot the production cross  section $\sigma$ of the LFV
process $e^{+}e^{-}\rightarrow\mu\tau$ as a function of $M_{Z}$
for three values of the parameter $k_{1}$: $k_{1}=0.2$(solid
line), 0.6(dashed line), and 1.0(dotted line). We can see from
Fig.1 that the production cross section $\sigma$ increases as
$k_{1}$ decreasing and strongly suppressed by large $M_{Z}$. For
$k_{1}\geq 0.6$, $M_{Z}\geq 3TeV$, the value of $\sigma$ is
smaller than $5.8\times 10^{-3}fb$, which is too small to be
detected. However, for $k_{1}=0.2$, $1TeV\leq M_{Z}\leq 3TeV$, the
cross section $\sigma$ varies in the range of $1.8\times
10^{-2}fb\sim 9.5fb$. In our calculation, we have assumed
$k_{\tau\mu}=\lambda=0.22$. If we let the flavor mixing factor
$k_{\tau\mu}$ increase, then the cross section $\sigma$ will be
enhanced. For example, if we take
$k_{\tau\mu}=\frac{1}{\sqrt{2}}$, the cross section $\sigma$ can
reach $97fb$ for $k_{1}=0.2$ and $M_{Z}=1TeV$.

TESLA is a proposed $e^{+}e^{-}$ linear collider with a center-of-mass
energy $\sqrt{s}=500GeV$ or $800GeV$. It is a multi-purpose machine that
will test various aspects of the SM and search for signals of NP beyond
the SM. It has a design luminosity of $5.8\times10^{34}cm^{-2}s^{-1}$
at $\sqrt{s}=800GeV$, which corresponds to $L=580fb^{-1}$ per year \cite{y14}.
Thus, for $k_{1}=0.2$, $1TeV \leq M_{Z}\leq 3TeV$, there will
be about 10 and up to 5533 events to be generated per year at
this machine.

A strong longitudinal polarization program at TESLA with
considerable polarization of the electron beam and the possibility
of polarization of the positron beam is planned \cite{y15}. Beam
polarization is not only useful for a possible reduction of the
background, but might also serve as a possible tool to disentangle
different contributions to the signal. Beam polarization of the
electron and positron beams would lead to a substantial
enhancement  of cross section of some processes. Considering the
polarization of electron and positron beams, the cross section
$\sigma$ of the LFV process $e^{+}e^{-}\rightarrow\mu\tau$ can be
written as:
\begin{equation}
\sigma=(1+P_{e})(1-P_{\bar{e}})(\sigma_{RR}+\sigma_{RL})
+(1-P_{e})(1+P_{\bar{e}})(\sigma_{LL}+\sigma_{LR}),
\end{equation}
where $P_{e}$ and $P_{\bar{e}}$ are the degrees of longitudinal electron
and positron polarization, respectively. $\sigma_{ij}$ are the chiral cross
section of this process, which can be easily given from Eq.(1).
We calculate the cross section $\sigma$ for different beam polarizations.
We find that the value of $\sigma$ for $(P_{e},P_{\bar{e}})=(0.8,0.6)$
is equal to that of $(P_{e},P_{\bar{e}})=(-0.8,0.6)$.
For $(P_{e},P_{\bar{e}})=(0.8,-0.6)$, the cross section $\sigma$
is larger than that of $(P_{e},P_{\bar{e}})=(0,0)$ in all of the
parameter space. In Fig.2 we plot $\sigma$ as a function of $M_{Z}$
for $k_{1}=0.2$ and different beam polarizations, in which the solid line,
dashed line, dotted line and dotted-dashed line represent
$(P_{e}, P_{\bar{e}})=(0,0), (0.8, 0.6), (0.8, -0.6)$,
and  $(-0.8, -0.6)$, respectively. From Fig.2 we can see that $\sigma$
is  indeed sensitive to the polarization of electron and positron beams.
For $(P_{e}, P_{\bar{e}})=(0.8,-0.6)$ and $k_{1}=0.2$,
the cross section $\sigma$
varies in the range of $4.2\times 10^{-2}fb\sim 22fb$ for
$1TeV\leq M_{Z}\leq 3TeV$.

To see the effect of center-of-mass energy $\sqrt{s}$ on the cross
section $\sigma$, we show the dependence of $\sigma$ on $\sqrt{s}$
for $k_{1}=0.2$, $M_{Z}=1.5TeV$ and different beam polarizations in
Fig.3. We can see from Fig.3 that the cross section $\sigma$
increases with $\sqrt{s}$ increasing for $\sqrt{s}\leq M_{Z}$.
For  $\sqrt{s}=M_{Z}$, $\sigma$ reach the maximum value.
The maximum value of $\sigma$ can reach $464fb$.

Although the LFV signal is quite spectacular, it is not background-free
\cite{y16}. The leading SM background of the LFV process
$e^{+}e^{-}\rightarrow\mu\tau$ come from the Drell-Yan process
$e^{+}e^{-}\rightarrow Z(\gamma)\rightarrow\tau^{+}\tau^{-}
\rightarrow\mu\bar{\nu}_{\mu}\nu_{\tau}\tau$ and $W^{+}W^{-}$ pair
production process $e^{+}e^{-}\rightarrow W^{+}W^{-}
\rightarrow\mu\bar{\nu}_{\mu}\tau\bar{\nu}_{\tau}$. As numerical estimation, we
only calculate the cross sections of the SM background at tree-level and
do not apply cut on the final state. Certainly, cuts as applied in the
study of lepton production will enhance the $S/\sqrt{B}$,
 which is the
ratio signal over square root of the background. In Fig.4 we plot
$S/\sqrt{B}$ as a function of the $Z^{\prime}$ mass $M_{Z}$
for $(P_{e},P_{\bar{e}})=(0,0)$
and three values of the
parameter $k_{1}$: $k_{1}=0.2$, $0.6$ and $1.0$. In Fig.4 we have taken the
integrated luminosity $L=580fb^{-1}$ for $\sqrt{s}=800GeV$ and the branching
ratios $Br(\tau\rightarrow\mu\bar{\nu}_{\mu}\nu_{\tau})=
(17.4\pm0.06)\%$, $Br(W\rightarrow \mu\bar{\nu}_{\mu})=(10.54\pm0.16)\%$
and $Br(W\rightarrow\tau\bar{\nu}_{\tau})=(11.09\pm0.22)\%$ \cite{y17}.
 From Fig.4 we can see that, with reasonable values of the parameters, the
LFV signals of extra $U(1)$ gauge bosons $Z^{\prime}$ may be
detected in the future LC experiments, even if any cuts are not
applied and the electron beam and the positron beam are not
polarized. If we apply appropriate cuts on the SM background, the
cases where the ratio $S/\sqrt{B}$ is of order 1 or smaller should
clearly improve. For example, a cut on the angular distribution of
the final state leptons will strongly reduce the $WW$
background\cite{y18}.

The results of the atmospheric neutrino experiment and the solar
neutrino experiment imply that large mixing is expected between
the second and third generation for leptons. In this paper, we
assume that the large mixing between the third and second
generation leptons is due to the new strong interactions and
consider the LFV process $e^{+}e^{-}\rightarrow \mu\tau$. We find
that the extra $U(1)$ gauge bosons $Z'$ predicted by TC2 models or
flavor-universal TC2 models can give significant contributions to
this process.  Over a sizable region of the parameter space, the
production cross  section $\sigma$ of this process is large than
$1.0\times 10^{-2}fb$. Furthermore, the cross section $\sigma$ can
be significantly enhanced by polarization of the electron beam and
the positron beam. For example, the value of $\sigma$ can reach
$22fb$ for $(P_{e},P_{\bar{e}})=(0.8,-0.6)$. Thus we can conclude
that the LFV signals of extra $U(1)$ gauge bosons $Z^{\prime}$ may
be detected via this LFV process $e^{+}e^{-}\rightarrow\mu\tau$ in
the future LC experiments.

\newpage
\begin{center}
{\bf Figure captions}
\end{center}

\begin{description}
\item[Fig.1:]The cross section $\sigma$ of the LFV process $e^{+}e^{-}
             \rightarrow\mu\tau$ as a function of the gauge boson $Z'$
             mass $M_{Z}$ for $\sqrt{s}=800GeV$, $k_{1}=0.2$(solid line),
$0.6$(dashed line) and $1.0$(dotted line).
\item[Fig.2:]The cross section $\sigma$ as a function of $M_{Z}$ for $\sqrt{s}=800GeV$,
             $k_{1}=0.2$ and different beam polarizations, in which the
             solid line, dashed line, dotted line and dotted-dashed line
             represent $(P_{e}, P_{\bar{e}})=(0,0)$, $(0.8,0.6)$,
              $(0.8, -0.6)$ and $(-0.8, -0.6)$, respectively.
\item[Fig.3:]The dependence of $\sigma$ on $\sqrt{s}$ for $k_{1}=0.2$,
             $M_{Z}=1.5TeV$ and different beam polarizations, in which
             the   solid line, dashed line, dotted line and dotted-dashed line
             represent $(P_{e}, P_{\bar{e}})=(0,0)$, $(0.8, 0.6)$,
             $(0.8, -0.6)$ and $(-0.8, -0.6)$, respectively.
\item[Fig.4:]The ratio of $S/\sqrt{B}$  as a function of $M_{Z}$ for  $L=580fb^{-1}$.
             and $k_{1}=0.2$(solid line), $0.6$(dashed line) and
             $1.0$(dotted line).
\end{description}

\newpage
\begin{figure}[htb]
\begin{center}
\begin{picture}(250,200)(0,0)
\put(-50,0){\epsfxsize120mm\epsfbox{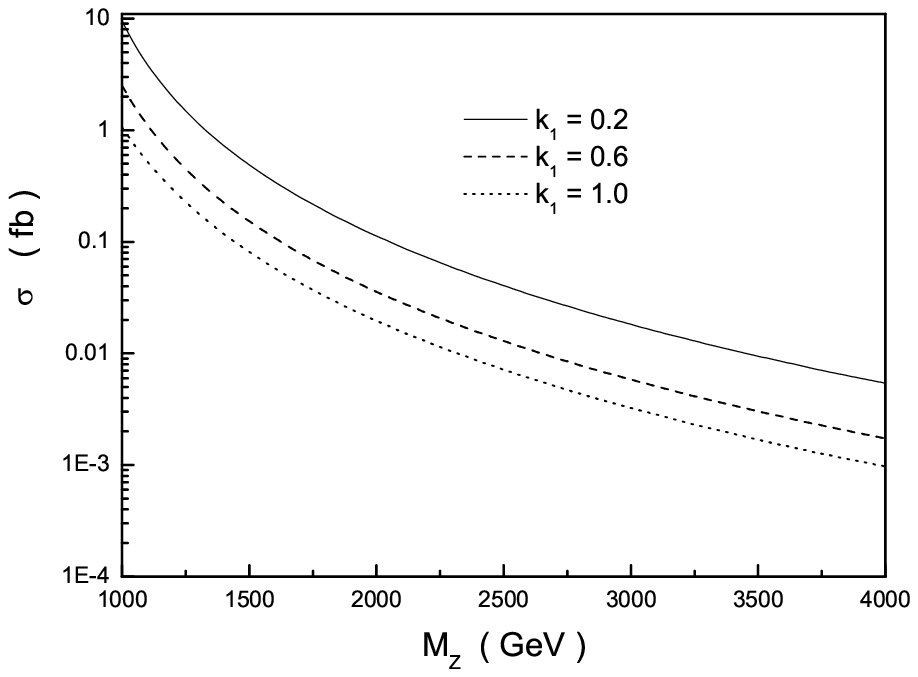}}
 \put(120,0){Fig.1}
\end{picture}
\end{center}
\end{figure}

\begin{figure}[htb]
\begin{center}
\begin{picture}(250,200)(0,0)
\put(-50,-50){\epsfxsize 120mm \epsfbox{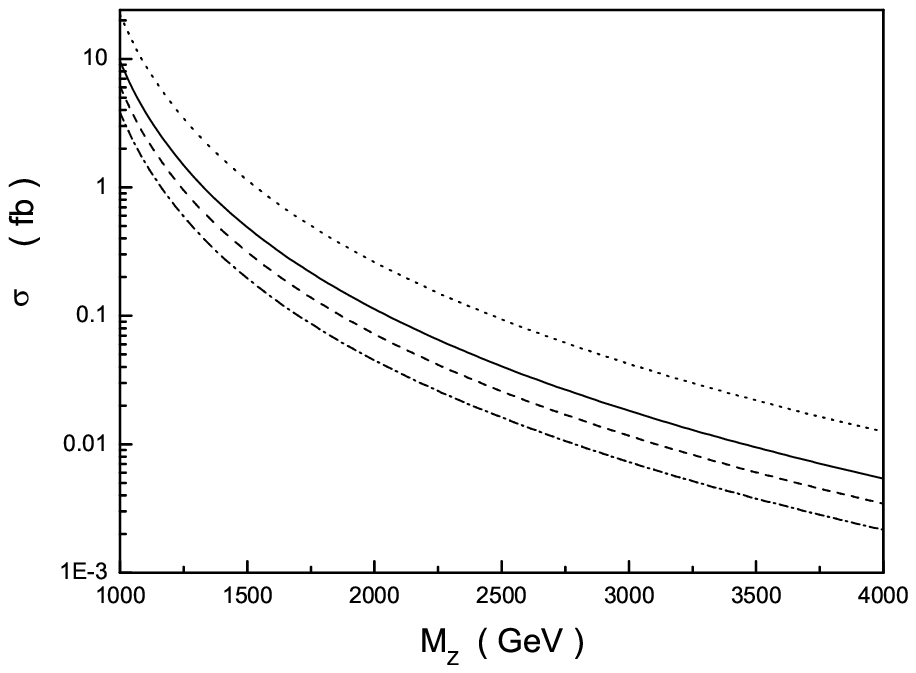}}
\put(120,-50){Fig.2}
\end{picture}
\end{center}
\end{figure}

\begin{figure}[hb]
\begin{center}
\begin{picture}(250,200)(0,0)
\put(-50,0){\epsfxsize 30mm\epsfbox{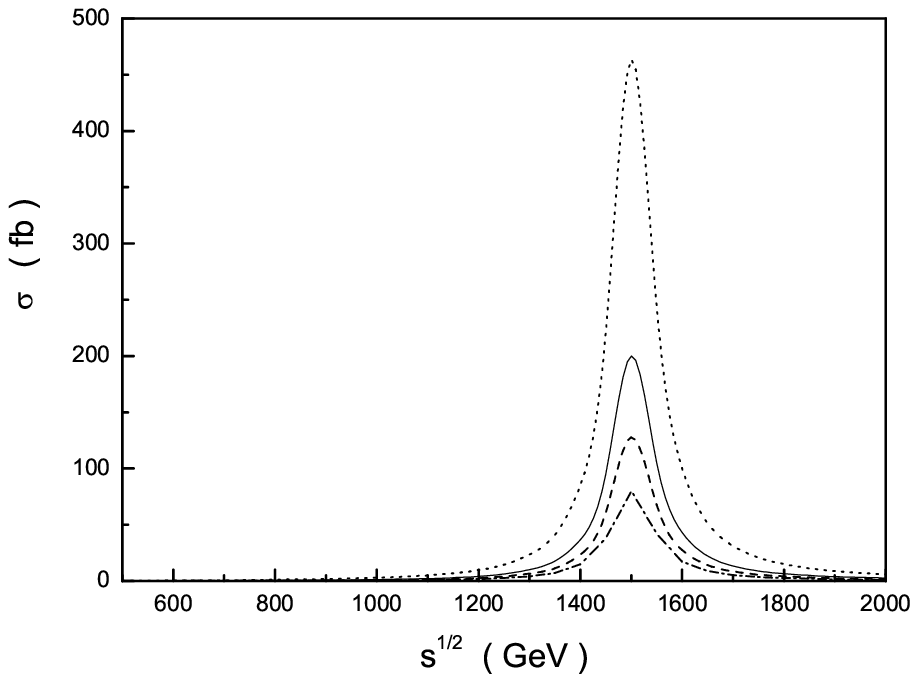}}
\put(120,0){Fig.3}
\end{picture}
\end{center}
\end{figure}

\begin{figure}[htb]
\begin{center}
\begin{picture}(250,200)(0,0)
\put(-50,50){\epsfxsize 120mm \epsfbox{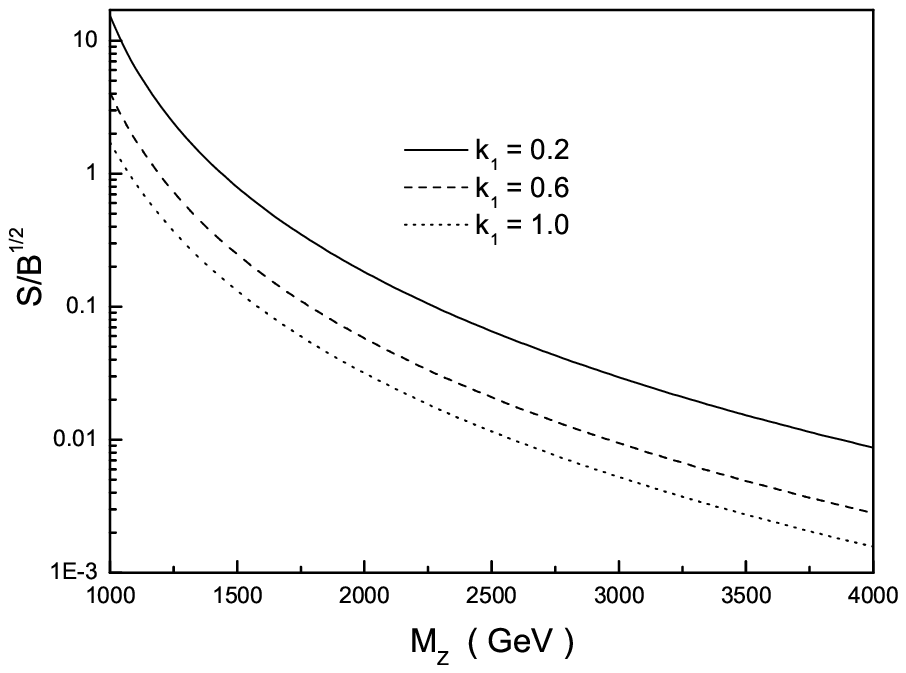}}
\put(120,30){Fig.4}
\end{picture}
\end{center}
\end{figure}

\end{document}